\documentclass[notitlepage,reprint,twocolumn,amsmath,amssymb,aps,prd,nofootinbib]{revtex4-1}

\usepackage{graphicx}
\usepackage{amsmath}
\usepackage{amssymb}
\usepackage{slashed}
\usepackage{physics}
\usepackage{subcaption}
\usepackage{xcolor}
\usepackage{hyperref}

\newcommand{\GeV}{{\rm GeV}}

\begin{document}
\title{A viable $L_e-L_\mu$ model with $\mu\to e$ violation }

\author{Marco Ardu}
\email{E-mail address: marco.ardu@umontpellier.fr}
\affiliation{LUPM, CNRS,
Université Montpellier
Place Eugene Bataillon, F-34095 Montpellier, Cedex 5, France}

\author{Fiona Kirk}
\email{E-mail address: fiona.kirk@itp.uni-hannover.de}
\affiliation{Physik-Institut, Universit\"at Z\"urich, Winterthurerstrasse 190, CH--8057 Z\"urich, Switzerland\\
Paul Scherrer Institut, CH--5232 Villigen PSI, Switzerland
}

\begin{abstract}
\vskip .5cm
\noindent
We extend the Standard Model gauge group by $U(1)_{L_e-L_\mu}$ and introduce two scalars, a doublet and a singlet, that are charged under this new group and have lepton flavour violating couplings. Since in this model $\mu \to e$ processes can only be mediated by $\mu\to \tau\times \tau\to e$ interactions, bounds from $\mu\to e$ transitions can be avoided while allowing for accessible new physics. We consider the case of a $Z'$ boson with a mass of $M_{Z'}\simeq10~$GeV and a gauge coupling $g'\simeq 10^{-4}$, which is in reach of Belle-II, and a long-lived $Z'$ boson with a mass of $\text{MeV}\lesssim M_{Z'}\lesssim m_\mu-m_e$ which can be probed by searching for $\ell\to \ell'+\text{inv.}$. Neutrino masses and mixing angles can also be accounted for if sterile neutrinos are added to the spectrum.

\end{abstract}


\maketitle

% \thanks{e-mail: marco.ardu@umontpellier.fr}
% \thanks{e-mail: fiona.kirk@psi.ch}


%\institute{LUPM, CNRS, Universit\' e Montpellier Place Eugene Bataillon, F-34095 Montpellier, Cedex 5, France
%           Physik-Institut, Universit\"at Z\"urich, Winterthurerstrasse 190, CH--8057 Z\"urich, Switzerland
%           Paul Scherrer Institut, CH--5232 Villigen PSI, Switzerland 
%}


\section{Introduction}
\label{sec:Introduction}
The success of the Standard Model of particle physics (SM) in describing the experimental data suggests that New Physics (NP) is either heavy or weakly coupled, or that it preferentially interacts with those sectors of the SM that are least constrained by experiment (e.g. the third generation of matter fields, in the case of Minimal Flavour Violation \cite{DAmbrosio:2002vsn}).

Processes that are suppressed or forbidden in the SM, such as lepton flavour violation (LFV) in the charged sector, are smoking gun signals of NP and constitute valuable probes for Beyond Standard Model scenarios.
Charged LFV has not yet been observed but is to be expected, since the discovery of neutrino masses and neutrino oscillations represents direct evidence for lepton flavour changing transitions. 
Searches for charged LFV can provide complementary tests of neutrino mass models, however, charged LFV is also predicted by popular BSM scenarios such as supersymmetry, which are motivated independently of neutrino masses~\cite{Ardu:2022sbt}.

The current upper limits on the rates of rare muon processes $\mathrm{Br}(\mu\to e\gamma)< 4\times 10^{-13}$ \cite{MEG:2016leq}, $\mathrm{Br}(\mu\to 3e)<10^{-12}$ \cite{SINDRUM:1987nra} and  $\mathrm{Br}(\mu A\to e A)< 7\times 10^{-13}$ ($\mu\to e$ conversion in nuclei) \cite{SINDRUMII:2006dvw} provide the most stringent constraints in models that allow for $\mu\to e$  transitions. 
The next generation of $\mu\to e$ experiments promise an improvement in sensitivity of up to four order of magnitude \cite{MEGII:2018kmf,Blondel:2013ia,COMET:2009qeh}. 
The bounds on LFV processes involving $\tau$ leptons are less constraining, with $\mathrm{Br}(\tau\to \ell \gamma), \mathrm{Br}(\tau\to 3\ell)\lesssim \mathrm{few}\times 10^{-8}, \ell = e,\mu$ \cite{BaBar:2009hkt,Hayasaka:2010np}. Here the experimental sensitivities are expected to improve by a factor $\sim 10$ in the near future \cite{Belle-II:2018jsg}.

If both $\tau\leftrightarrow e$ and $\tau \leftrightarrow \mu$ couplings are present, there is no symmetry forbidding $\mu\leftrightarrow e$ processes mediated by the product of $\mu\to \tau$ and $\tau\to e$ interactions. Considering that the future sensitivities of $\mu\leftrightarrow e$ and $\tau\leftrightarrow \ell$ searches approximately satisfy the inequality
\begin{equation}
     \mathrm{Br}(\mu \to e)\lesssim \mathrm{Br}(\tau\to \mu)\mathrm{Br}(\tau \to e), \label{eq:mutoevstaul}
\end{equation}
$\mu\to e$ observables can probe products of $\mu\leftrightarrow \tau$ and $\tau\leftrightarrow e$ couplings which are beyond the reach of direct $\tau\to \ell$ searches. If the NP is heavy, model independent $(\mu\to \tau)\times(\tau\to e)$ contributions can be calculated in the Standard Model Effective Field Theory (SMEFT)~\cite{Ardu:2022pzk}. Since these correspond to the combination of two dimension 6 operators, they only arise at dimension 8. In models, the combination of $\mu\to\tau$ and $\tau\to e$ can be larger than dimension 8 effects, making it more accessible to $\mu \to e$ experiments.

The aim of this article is to explore the sensitivity of $\mu\to e$ processes to $\mu \to \tau \times \tau \to e$ transitions in a UV-complete model with light new physics, which we cannot parametrise with the SMEFT.
A particularly simple way to extend the SM is to enlarge the SM gauge group by a new spontaneously broken abelian $U(1)'$ group. This leads to new interactions mediated by a neutral massive vector boson, commonly referred to as a $Z'$ gauge boson.
$U(1)'$ groups may be remnants of larger non-abelian groups, such as $SO(10)$ or $E_6$, which feature in certain Grand Unified Theories (GUTs) \cite{Langacker:2008yv} or they can be considered as standalone extensions, as in the case of $U(1)_{L_e-L_\mu}$, $U(1)_{L_\mu-L_\tau}$,  $U(1)_{L_e-L_\tau}$ \cite{He:1991qd} or $U(1)_{B-L}$ \cite{CARLSON1987378}, which are anomaly-free global symmetries of the SM.

LFV $Z'$ couplings are introduced in such models if the mass and gauge eigenstates of the leptons are misaligned.
The stringent bounds from $\mu\to e$ transitions \cite{Langacker:2000ju,Murakami:2001cs,Chiang:2011cv} can be avoided in $U(1)'$ models that allow only for $\tau\leftrightarrow \ell$,  flavour changes, $\ell=e,\mu$~\cite{Heeck:2011wj,Heeck:2014qea,Heeck:2016xkh}, which are less constrained. 

We propose a $Z'$ model where $\mu\to e$ transitions are mediated by the product of $\mu\leftrightarrow \tau$ and $\tau \leftrightarrow e$ couplings only. In this way we suppress $\mu \to e$ rates enough to respect the current upper limits, while being in reach of future $\mu \to e$ experiments. We gauge $L_e-L_\mu$\footnote{In models with gauged $L_{\mu}-L_{\tau}$ or $L_{e}-L_\tau$, where both $\mu\leftrightarrow \tau$ and $e\leftrightarrow\tau$ couplings are present, gauge invariance does not forbid $\mu\leftrightarrow e$ couplings.}, aiming for a feebly-coupled $Z'$ with a mass below the electroweak scale,
and extend the scalar sector by an extra doublet and a new singlet, both of which are charged under the new $U(1)'$. At some unknown high energy, the vacuum expectation value (VEV) of the singlet, $v_S$,  breaks $U(1)_{{L_e}-L_\mu}$, gives the $Z'$ boson a mass and generates Majorana masses for the sterile neutrinos that are charged under $U(1)_{{L_e}-L_\mu}$, setting the stage for a type I seesaw neutrino mass model.
LFV is introduced via the Yukawa interactions with the new scalars, and the $Z'$ boson receives flavour changing couplings when the $U(1)'$ charged doublet acquires a non-vanishing vacuum expectation value (VEV). 

A large region of the model's parameter space is in reach of future $Z'$ and LFV searches and will be tested in the upcoming years.

This article is organized as follows: In Section \ref{sec:themodel} we present the particle content of the model and define our notation. In Section \ref{sec:Pheno} the phenomenology is studied. We briefly review the constraints on non-SM gauge interactions of electrons and muons and give the rates of $\tau\to \ell$ and $\mu\to e$ processes.    

\section{The Model}
\label{sec:themodel}

We extend the Standard Model gauge group $SU(3)_c\times SU(2)_L\times U(1)_Y$ by the abelian anomaly-free $U(1)_{L_1-L_2}\equiv U(1)'$ and consider the particle content summarised in Table \ref{tab:Fields}.
\begin{table}[t]
\begin{center}
\begin{tabular}{c c c | c }
\hline\noalign{\smallskip}
                                  &                    &                      & $U(1)'$\\
\noalign{\smallskip}\hline\noalign{\smallskip}
$L_1$    (1,\;2,\;$-\frac{1}{2}$) & $e_1$   (1,\;1,\;-1) & $N_1$    (1,\;1,\;0) & +1     \\
$L_2$  (1,\;2,\;$-\frac{1}{2}$) & $e_2$ (1,\;1,\;-1) & $N_2$  (1,\;1,\;0) & $-$1     \\
$L_3$ (1,\;2,\;$-\frac{1}{2}$) & $e_3$(1,\;1,\;-1) & $N_3$ (1,\;1,\;0) & 0      \\
\noalign{\smallskip}\hline\noalign{\smallskip}
$H$      (1,\;2,\;$\frac{1}{2}$)  &                    &                      & 0      \\     
$\phi$   (1,\;2,\;$\frac{1}{2}$)  &                    &                      & $-$1     \\
$S$      (1,\;1,\;0)              &                    &                      & $-$1     \\
\noalign{\smallskip}\hline
\end{tabular}
\caption{Field content of the model. In parentheses we give the representations under the SM gauge group ($SU(3)_c$, $SU(2)_L$, $U(1)_Y$), in the last column the charges under $U(1)'$.} \label{tab:Fields}
\end{center}
\end{table}

Then the most generic renormalisable Lagrangian is 
\begin{equation}
    \mathcal{L}=\mathcal{L}_{\rm kin}-\mathcal{L}_{\rm Yuk}-V(H,\phi,S)
\end{equation}
where
\begin{align}
    \mathcal{L}_{\rm kin}=&-\frac{1}{4}B^{\alpha\beta}B_{\alpha\beta}-\frac{1}{4}W^{a\alpha\beta}W^a_{\alpha\beta}-\frac{1}{4}B'^{\alpha\beta}B'_{\alpha\beta}\nonumber\\
        &+\sum_{\Phi=H, \phi, S}(D_\alpha \Phi)(D^\alpha \Phi)^\dagger+\sum_{\psi} i\overline{\psi} \slashed{D}\psi,\label{eq:Lkin}\\
    \mathcal{L}_{\rm Yuk}&=\sum_{i\in \lbrace 1,2,3\rbrace}y_{ii}\bar{L}_i e_i H + y^\nu_{ii} \bar{L}_i N_i \tilde{H} +\mathrm{h.c.}\nonumber\\
&+ y_{3 1}\bar{L}_3 e_1\phi + y_{23}\bar{L}_2 e_3 \phi\nonumber\\
&+ y_{13}^\nu \bar{L}_1 N_3 \tilde{\phi} + y_{3 2}^\nu \bar{L}_3 N_2 \tilde{\phi} + \frac{1}{2} y^\nu_{13} \bar{N}_1^c N_3 S\nonumber\\
&+\frac{1}{2} M^N_{33} \bar{N}_3^c N_3 + \frac{1}{2} M^N_{12} \bar{N}_1^c N_2 + \mathrm{h.c.},\label{eq:LYuk}\\
 V(H,\phi, S)&= m_{H}^2 H^\dagger H + m_{\phi}^2 \phi^\dagger \phi - m_{S}^2 S^\dagger S\notag\\
 & - m_{\phi H S} \left(\left(\phi^\dagger H\right) S+ S^\dagger \left(H^\dagger \phi\right)\right)\notag\\
 & +\frac{1}{2}\lambda_{H}\left(H^\dagger H\right)^2 + \frac{1}{2}\lambda_{\phi}\left(\phi^\dagger \phi\right)^2 
 + \frac{1}{2}\lambda_S\left(S^\dagger S\right)^2 \notag\\
 & + \lambda_{H\phi} \left(H^\dagger H\right)\left(\phi^\dagger \phi\right)
 + \tilde{\lambda}_{H\phi} \left(H^\dagger \phi\right)\left(\phi^\dagger H\right)\notag\\
 & + \lambda_{HS} \left(H^\dagger H\right)\left(S^\dagger S\right)  + \lambda_{\phi S} \left(\phi^\dagger \phi\right)\left(S^\dagger S\right)\,\label{eq:Lagrangian}.
\end{align}
Disregarding $SU(3)_c$, we use the following conventions for the covariant derivative
\begin{align*}
    D_\mu=\partial +ig_1YB_\mu+ig_2 \frac{\tau^a}{2} W_\mu^a+ig' Q'B_\mu'  \,,
\end{align*}
where $g_1, g_2 $ and $g'$ are the gauge couplings of $U(1)_Y$, $SU(2)_L$ and $U(1)'$, respectively, while $\tau^a$ are the Pauli matrices. The label $\psi$ in Eq.~\eqref{eq:Lkin} runs over all fermions of the model. We identify the gauge eigenstates $1,2,3$ by their diagonal Yukawa couplings with the $U(1)'$ neutral doublet $H$.

$U(1)'$ gauge invariance allows for off-diagonal Yukawa interactions of type $2\leftrightarrow3$ and $1\leftrightarrow3$ with the doublet $\phi$, but forbids interactions of type $1\leftrightarrow 2$ among the charged leptons.
The off-diagonal Yukawas $y_{31}, y_{23}$ are the only  parameters of the charged lepton sector that introduce LFV. 
As we discuss in Section \ref{ssec:intvsmassbasis}, the mass eigenstates $e,\mu,\tau$ are nearly aligned with the interaction eigenstates $1,2,3$, and as a result  $\mu\to e$ transitions are controlled by the product $y_{31}\times y_{23}$.
Due to the approximate alignment of gauge eigenstates and mass eigenstates, we will use the term \emph{lepton flavour} in both bases and refer to $U(1)_{L_1-L_2}$ as $U(1)_{L_e-L_\mu}$.

The scalar potential parameters can be such that all scalars acquire VEVs, $\expval{H}=v_H/\sqrt{2}$, $\expval{\phi}=v_\phi/\sqrt{2}$,  $\expval{S}=v_S/\sqrt{2}$.
$v_H$ and $v_\phi$  spontaneously break the electroweak gauge symmetry and thus must satisfy
\begin{align*}
    \sqrt{v_H^2+v_\phi^2}=v \qquad \mathrm{with} ~ v=246 ~ \GeV\,,
\end{align*}
whereas either $v_S$ or $v_\phi$ can be the VEV that breaks $U(1)'$ and provides the dominant contribution to the $Z'$ mass. Since we are gauging the lepton flavor difference $L_e-L_\mu$, resulting in a $Z'$ boson that couples to electrons and muons,
several experiments can constrain the gauge coupling $g'$. Values larger than $g'\sim \mathrm{few}\times10^{-4}$ are excluded in a vast region of the $Z'$ mass vs. $g'$ coupling plane. Since, as discussed in Section~\ref{sec:Pheno}, we are aiming at a $Z'$ mass in the GeV range, we will focus on the limit $v_S\gg v$.

\subsection{Lepton Mass Basis}
\label{ssec:intvsmassbasis}

Upon spontaneous symmetry breaking, the Yukawa interactions contribute  as follows to the mass matrix of the charged leptons:
\begin{align*}
&\sum_{i,j\in \lbrace 1,2, 3\rbrace} \bar{L}_i \mathcal{M}_{ij} e_j + \mathrm{h.c.}=\notag \\
& \quad
\begin{pmatrix}
\bar{L}_1 & \bar{L}_2 & \bar{L}_3
\end{pmatrix}
\begin{pmatrix}
\frac{v_H}{\sqrt{2}} y_{11}           &  0            &  0                     \\
0                  & \frac{v_H}{\sqrt{2}} y_{22}  &  \frac{v_\phi}{\sqrt{2}} y_{2 3}   \\
\frac{v_\phi}{\sqrt{2}} y_{3 1}  &  0            &  \frac{v_H}{\sqrt{2}} y_{33}
\end{pmatrix}
\begin{pmatrix}
e_1\\
e_2\\
e_3
\end{pmatrix}
+ \mathrm{h.c.}
\end{align*}
Without loss of generality, we can take the Yukawa couplings to be real, as it is possible to absorb the complex phases in the field definitions. As a result, the symmetric matrices $\mathcal{MM}^T$ and $\mathcal{M}^T\mathcal{M}$ can be diagonalised by the orthogonal matrices $O^L$ and $O^R$, respectively,
\begin{align*}
&(O^L)^T\mathcal{M}\mathcal{M}^{ T} O^L=\mathcal{M}_{\rm diag}^2\,,\\
&(O^R)^T\mathcal{M}^{T} \mathcal{M} O^R=\mathcal{M}_{\rm diag}^2\, ,
\end{align*}
where $\mathcal{M}_{\rm diag}^2=\mathrm{diag}\begin{pmatrix}
m_e^2 & m^2_\mu & m^2_\tau
\end{pmatrix}$ has non-negative diagonal entries corresponding to the squared charged lepton masses.

Assuming the flavour off-diagonal Yukawas to be smaller than the flavour conserving ones, the angles that rotate the gauge eigenbasis into the mass eigenbasis can be treated perturbatively, and the orthogonal matrices $O^{L,R}$ can be written as $O^{L,R}=R^{L,R}_{13} R^{L,R}_{23}$, where 
\begin{align*}
R_{13}^{L,R}=\begin{pmatrix}
1 & 0 & \theta_{13}^{L,R}\\
0 & 1 & 0\\
-\theta_{13}^{L,R} & 0 & 1
\end{pmatrix},
\quad
R^{L,R}_{23}=\begin{pmatrix}
1 & 0 & 0\\
0 & 1 & -\theta_{23}^{L,R}\\
0 & \theta_{23}^{L,R} & 1
\end{pmatrix}\,.
\end{align*}
At leading order in the ratio of flavour off-diagonal Yukawa couplings and flavour diagonal Yukawa couplings, the rotation angles read
\begin{align}
\theta_{13}^L&\simeq \frac{v_\phi}{v_H}\frac{y_{3 1}y_{11}}{y_{33}^2}\nonumber\\
\theta_{23}^L&\simeq -\frac{v_\phi}{v_H}\frac{y_{23}}{y_{33}}\nonumber\\
\theta_{13}^R&\simeq \frac{v_\phi}{v_H}\frac{y_{3 1}}{y_{33}}\nonumber \\
\theta_{23}^R&\simeq -\frac{v_\phi}{v_H}\frac{y_{23}y_{22}}{y_{33}^2} 
\label{eq:anglesmasslepton}
\end{align}
and the charged lepton masses are given by
\begin{align}
\mathrm{diag}\begin{pmatrix} m_e & m_\mu & m_\tau \end{pmatrix} \simeq \frac{v_H}{\sqrt{2}}\ \mathrm{diag}\begin{pmatrix}  y_{11} & y_{22} & y_{33} \end{pmatrix}\,\label{eq:leptmass},
\end{align}
where $v_H$ is the VEV of the $U(1)'$ neutral Higgs doublet. Note that the electron and muon masses receive contributions from the off-diagonal couplings of the form $\sim v_H y_{11} \theta^2,\ v_H y_{22}\theta^2$, hence it is sufficient that the angles $\theta$ are small to prevent large corrections to Eq.~(\ref{eq:leptmass}).

\subsection{Gauge Sector}

The mass matrix of the neutral gauge bosons can be obtained from the kinetic terms of the scalars in unitary gauge:\\
\begin{align*}
& \left(D_\alpha H^\dagger\right)\left(D^\alpha H\right)+ 
\left(D_\alpha \phi^\dagger\right)\left(D^\alpha \phi\right)+ 
\left(D_\alpha S^\dagger\right)\left(D^\alpha S\right)\supset\nonumber\\
&\frac{1}{8}
\begin{pmatrix}
B_\alpha \\ W_\alpha^3 \\ B_\alpha '
\end{pmatrix}^{\!\!T}\!\!
\begin{pmatrix}
g_1^2 v^2     & -g_1 g_2 v^2 & -2 g' g_1 v_\phi^2\\
-g_1 g_2 v^2  & g_2^2 v^2    & 2 g' g_2 v_\phi^2\\
-2 g' g_1 v_\phi^2                      & 2 g' g_2 v_\phi^2                      & 4 g^{\prime 2} \left(v_S^2 + v_\phi^2\right)
\end{pmatrix}
\begin{pmatrix}
B^\alpha\\
W^{3\alpha} \\
B'^\alpha
\end{pmatrix}
\end{align*}
with $v^2 \equiv v_H^2 + v_\phi^2$.
If $M_{Z'}\lesssim 10~$GeV, $g'$ must be $\lesssim 10^{-4}$ to avoid the constraints on electron fifth-force (see Section \ref{ssec:PhenononLFV}), leading to suppressed mixing, $\propto g'$, between the $B'$ boson and the SM gauge bosons. In the limit $v_S\gg v$, we approximate the mass matrix as follows:
\begin{align*}
\frac{1}{8}
\begin{pmatrix}
B_\alpha & W_\alpha^3 & B_\alpha '
\end{pmatrix}
\begin{pmatrix}
g_1^2 v^2     & -g_1 g_2 v^2 & 0\\
-g_1 g_2 v^2  & g_2^2 v^2   & 0\\
0                                       & 0                                      & 4 g^{\prime 2} v_S^2 
\end{pmatrix}
\begin{pmatrix}
B^\alpha\\
W^{3\alpha} \\
B'^\alpha
\end{pmatrix}\,.
\end{align*}
The mass eigenstate $Z'$, with $M_{Z'}=g'v_s$, is aligned with $B'$, while the photon $A$ and the $Z$ boson (with $M_Z=v/2\sqrt{g_1^2+g_2^2}$) are related to the interaction basis via a rotation by the Weinberg angle $\theta_W$, with $\tan\theta_W\equiv g_1/g_2$, just as in the SM.
\begin{align*}
\begin{pmatrix}
 A_\alpha\\
 Z_\alpha\\
 Z'_\alpha
\end{pmatrix}=\begin{pmatrix}
  \cos \theta_W & \sin \theta_W & 0\\
  -\sin \theta_W & \cos \theta_W & 0\\
  0 & 0 & 1
\end{pmatrix}\begin{pmatrix}
B_\alpha\\
W^{3}_\alpha \\
B'_\alpha
\end{pmatrix}
\end{align*}
 Note that kinetic mixing of the abelian fields, $\epsilon B^{\mu\nu} B'_{\mu\nu}$, is not forbidden by any symmetry and introduces a new coupling $\epsilon$ that leads to $g'$ independent interactions of the $Z'$ boson with the SM fermions. If the $U(1)'$ is the remnant of a larger spontaneously broken non-abelian gauge group, kinetic mixing is absent at tree-level. However, kinetic mixing can always be generated via loops involving fermions that are charged under both abelian groups. These lead to the finite and calculable contribution  $\epsilon_{1-\rm loop}\sim g' g_1/(16 \pi^2)\log(m_\mu/m_e)$. In order to obtain a more predictive model, we consider vanishing tree-level $\epsilon$ but loop-induced  kinetic mixing.

The lepton mass eigenstates are related to the gauge eigenstates via $L_i=O^L_{ i\alpha} L_\alpha$, $e_i=O^R_{i\alpha } \alpha$ with $\alpha\in \{e,\mu,\tau\}$ and $i\in\{1,2,3\} $. Whereas the flavour universal photon and $Z$ couplings are unaffected by these unitary transformations, the lepton flavour non-universal $Z'$ couplings are sensitive to the misalignment of the lepton gauge and mass eigenbases and receive flavour changing contributions. The $Z'$ interactions with the charged leptons take the form
\begin{align}
    &\mathcal{L}^{\rm LFV}_{Z'}= -g'Z'_\rho\bigg[ \nonumber\\
    &\quad \theta^L_{13}(\overline{L}_e \gamma^\rho L_\tau+\tau\leftrightarrow e)+\theta^L_{23}(\overline{L}_\mu \gamma^\rho L_\tau+\tau\leftrightarrow \mu) \nonumber\\
    &\quad+\theta^L_{13}\theta^L_{23}(\overline{L}_e \gamma^\rho L_\mu+\mu\leftrightarrow e)+\theta^R_{13}(\overline{e} \gamma^\rho \tau+\tau\leftrightarrow e)\nonumber\\
    &\quad +\theta^R_{23}(\overline{\mu} \gamma^\rho \tau+\tau\leftrightarrow \mu)+\theta^R_{13}\theta^R_{23}(\overline{e} \gamma^\rho \mu+\mu\leftrightarrow e)\bigg] \label{eq:ZprimeLFV}
\end{align}
where $\theta^{L,R}$ are the rotation angles given in Eq.~(\ref{eq:anglesmasslepton}). The $\mu\to e$ couplings are proportional to the product of the off-diagonal $y_{31}\times y_{23}$ Yukawas that parametrise the $\tau \leftrightarrow e$ and $\tau \leftrightarrow \mu$ flavour changes. Note that the products $\theta_{13}^L \theta_{23}^L$ and $\theta_{13}^R \theta_{23}^R$ in the tree-level $\mu\to e$ $Z'$ interactions are suppressed by the SM electron and muon Yukawas respectively. Only the combination $\theta_{13}^R \theta_{23}^L$ is sizeable.

\subsection{Scalar Sector}
\label{ssec:scalars}

We supplement the SM Higgs sector by two new scalars, an $SU(2)_L$ doublet $\phi$ and a singlet $S$, both of which are charged under $U(1)'$. Assuming the singlet to be heavier than the doublet, we integrate it out, reducing the scalar sector to that of a regular two Higgs Doublet Model (2HDM) with the doublets $H$ and $\phi$. Since the VEV of the singlet breaks $U(1)'$, the scalar potential now features  all $SU(2)_L\otimes U(1)_Y$ gauge invariant potential terms, including $U(1)'$ breaking interactions, which are generated via singlet VEV insertions:
\begin{align}
  \mathcal{V}^{\mathrm{eff}}=
 & M_{HH}^2 H^\dagger H + M_{\phi\phi}^2 \phi^\dagger \phi
  -M_{H\phi}^2\left(H^\dagger \phi+\phi^\dagger H\right)\notag\\
 & +\frac{1}{2}\Lambda_{H}\left(H^\dagger H\right)^2 + \frac{1}{2}\Lambda_{\phi}\left(\phi^\dagger \phi\right)^2 \notag\\
 & + \Lambda_{H\phi} \left(H^\dagger H\right)\left(\phi^\dagger \phi\right)
 + \tilde{\Lambda}_{H\phi} \left(H^\dagger \phi\right)\left(\phi^\dagger H\right)\notag\\
 & + \frac{1}{2}\Lambda_a \left(\left(H^\dagger \phi\right)^2 + \left(\phi^\dagger H\right)^2\right)\notag\\
 & + \Lambda_b \left(H^\dagger H\right)\left(\left(H^\dagger \phi\right)+\left(\phi^\dagger H\right)\right)\notag\\
 & + \Lambda_c \left(\phi^\dagger \phi\right)\left(\left(H^\dagger \phi\right)+\left(\phi^\dagger H\right)\right)\,.
 \label{eq:EffScalarPotential}
\end{align}
It is particularly convenient to rotate into the so-called Higgs basis, where only one of the doublets acquires a vacuum expectation value:
\begin{align*}
    \begin{pmatrix}
    H_1\\
    H_2 
    \end{pmatrix}=\begin{pmatrix}
      \cos\beta & \sin\beta \\
      -\sin\beta & \cos\beta 
    \end{pmatrix}\begin{pmatrix}
      H\\
      \phi
    \end{pmatrix}\qquad \mathrm{with}\ \tan \beta\equiv v_\phi/v_H
\end{align*}
$\expval{H_1}=v/\sqrt{2}$, with $v=\sqrt{v_\phi^2+v_H^2}$, and $\expval{H_2}=0$. We relabel the potential parameters as follows
\begin{align}
	V(H_1,H_2)
	&=m^2_{11}H^\dagger_{1}H_1+m^2_{22}H^\dagger_{2}H_2-m^2_{12}(H^\dagger_1H_2+\mathrm{h.c})\nonumber\\
	&+\frac{\lambda_1}{2}(H^\dagger_1 H_1)^2+\frac{\lambda_2}{2}(H^\dagger_2 H_2)^2\nonumber \\
	&+\lambda_3 (H^\dagger_1 H_1)(H^\dagger_2 H_2)+\lambda_4(H^\dagger_1 H_2)(H^\dagger_2 H_1)\nonumber\\
	&+\left(\frac{\lambda_5}{2}(H^\dagger_1 H_2)^2
	+\lambda_6(H^\dagger_1 H_1)(H_1H^\dagger_2)\right.\nonumber\\
	& \left.\qquad+\lambda_7(H^\dagger_2 H_2)(H_1H^\dagger_2)+\mathrm{h.c}\right), \label{eq:potentialHiggsbasis}
\end{align}
and expand the doublets in terms of the canonically normalized scalar fields
\begin{align*}
	H_1=\begin{pmatrix}
		G^{+} \\
		\frac{1}{\sqrt{2}}(v+H^0_1+iG^0)
	\end{pmatrix}\,,
	\quad 
	H_2=\begin{pmatrix}
	H^{+} \\
	\frac{1}{\sqrt{2}}(H^0_2+iA)
\end{pmatrix}\,.
\end{align*} 
Once the Goldstone bosons $G$ are eaten by the electroweak gauge bosons, the spectrum contains one charged scalar $H^+$, two CP even scalars $H^0_1,H^0_2$ and one CP odd neutral scalar $A$. If the potential parameters are real, the physical states are CP eigenstates and only the two CP even neutral scalars, $H_1^0$ and $H_2^0$, can mix. The mass matrix of $H_1^0$ and $H_2^0$ is diagonalised by the angle $\beta-\alpha$.
We identify two scalar mass-eigenstates
\begin{align*}
    h= \sin(\beta-\alpha) H^0_1 + \cos(\beta-\alpha)H^0_2\\
    \rho=\cos(\beta-\alpha)H^0_1-\sin(\beta-\alpha) H^0_2 
\end{align*}
with the masses $m_h$ and $m_{\rho}$, respectively.
The angle $\beta-\alpha$ can be written in terms of $\lambda_6$, which is defined in Eq.~(\ref{eq:potentialHiggsbasis}), as well as the scalar masses $m_h$ and $m_\rho$~\cite{Davidson:2005cw,Davidson:2016utf}:
\begin{equation}
	\cos(\beta-\alpha)\sin(\beta-\alpha)\equiv c_{\beta \alpha}s_{\beta\alpha}=-\frac{\lambda_6v^2}{(m^2_{\rho}-m_h^2)}
	\label{eq:Higgsbasisangle}
\end{equation}
The decoupling limit of the 2HDM is obtained when the mass term of the VEV-less doublet $H_2$ satisfies the condition $m_{22}\equiv M\gg v$. The doublet mass terms in the Higgs basis, $m_{11}$, $m_{12}$, $m_{22}$ (see Eq.~\eqref{eq:potentialHiggsbasis}), are related to the mass matrix in the $H,\phi$ basis (see Eq.~\eqref{eq:EffScalarPotential}) in the following way
\begin{equation}
    m^2_{11}=c_\beta^2\left(M^2_{HH}+M^2_{\phi\phi} t_\beta^2-M^2_{H\phi}\frac{2 t_\beta}{\sqrt{1+t_\beta^2}} \right)
    \label{eq:m11}
\end{equation}
\begin{equation}
    m^2_{22}=c_\beta^2\left(M^2_{\phi\phi}+M^2_{H H} t_\beta^2+M^2_{H\phi}\frac{2 t_\beta}{\sqrt{1+t_\beta^2}} \right)
\end{equation}
\begin{equation}
    m^2_{12}=\frac{1}{2}\left(M^2_{HH}-M^2_{\phi\phi}\right)s_{2\beta}+M^2_{H\phi}c_{2\beta} \label{eq:m12}.
\end{equation}
The potential minimum conditions imply
\begin{equation}
    m^2_{11}=-\frac{1}{2}\lambda_1 v^2\qquad m^2_{12}=\frac{1}{2}\lambda_6 v^2\,.
\end{equation}
Consequently, if we assume $\lambda_1$ and $\lambda_6$ to be perturbative couplings, $m_{11}$ and $m_{12}$ are sub-electroweak masses.
The decoupling condition $m_{22}\gg v$ is satisfied if $M_{HH},M_{\phi\phi},$ $M_{H\phi}\gg v$, however, in this case the minimum potential conditions require a fine-tuned cancellation in Eqs.~\eqref{eq:m11} and \eqref{eq:m12}. If, on the other hand, $\beta \ll 1$, the mass configuration $M_{H}\sim v$ and $M_{\phi\phi},M_{H\phi}\gg v$ leads to natural minimum potential conditions in the decoupling limit. To avoid fine-tuned spontaneous symmetry breaking, we consider $\beta\ll 1$.

Once $H_2$ is decoupled, we identify the light CP even scalar $h$ as the 125-\GeV\ Higgs boson, while the orthogonal state $\rho$ has a mass $m_{\rho}\sim  m_{22}\sim M$. In the decoupling limit, the relation~\cite{Davidson:2005cw}
\begin{align*}
    m_{\rho}^2-m^2_{A}=(\lambda_5+\lambda_1) v^2-m_h^2,
\end{align*}
implies that the pseudoscalar $A$ and the scalar $\rho$ are approximately degenerate, with a mass-splitting of order $v^2$.

In the lepton mass basis, the doublet $H_1$, which is aligned with the electroweak vacuum, has diagonal Yukawa interactions with the leptons, while the $H_2$ couplings are in general flavour-changing. Written in terms of the Yukawa couplings defined in Eq.~(\ref{eq:LYuk}), the lepton Yukawa sector in the Higgs basis reads
\begin{align*}
    \mathcal{L}_{\rm Yuk}=\frac{\sqrt{2} \delta_{ij} m_{i}}{v}\bar{L}_i e_i H_1 + (O^T_{L} Y_2 O_R)_{ij}\bar{L}_i e_j H_2 +\mathrm{h.c}\,,
\end{align*}
where the sum over $i,j\in \lbrace e,\mu,\tau\rbrace$ is understood and 
\begin{align*}
    Y_2=\begin{pmatrix}
     -y_{11} \sin\beta & 0 & 0 \\
     0 & -y_{22} \sin\beta & y_{23}\cos\beta \\
     y_{31}\cos\beta & 0 & -y_{33}\sin\beta 
    \end{pmatrix}.
\end{align*}
Defining $\mathcal{Y}\equiv O^T_{L} Y_2 O_R $, the couplings of the neutral scalar mass eigenstates to the leptons take the form
\begin{align}
  \mathcal{L}^{\rm neut}_{\rm Yuk}=%\sum_{i,j\in \lbrace e,\mu,\tau\rbrace}
  &\frac{h}{\sqrt{2}}\bar{e}_i\left[s_{\beta\alpha}\frac{\sqrt{2} \delta_{ij} m_{i}}{v}+c_{\beta\alpha}(\mathcal{Y}_{ij}P_R+\mathcal{Y}^\dagger_{ij}P_L)  \right]e_j\nonumber\\
    +&\frac{\rho}{\sqrt{2}}\bar{e}_i\left[c_{\beta\alpha}\frac{\sqrt{2} \delta_{ij} m_{i}}{v}-s_{\beta\alpha}(\mathcal{Y}_{ij}P_R+\mathcal{Y}^\dagger_{ij}P_L)  \right]e_j\nonumber\\
    +&\frac{iA}{\sqrt{2}}\bar{e}_i\left[\mathcal{Y}_{ij}P_R-\mathcal{Y}^\dagger_{ij}P_L  \right]e_j\,. \label{eq:scalarYuk}
\end{align}
Since the rate of $h\to \tau^+\tau^-$ measured at the LHC \cite{ATLAS:2018ynr,CMS:2017zyp} is compatible with the Standard Model prediction, we require $s_{\beta\alpha}\sim 1$. As a result, the flavour-changing couplings of the Higgs boson $h$ are suppressed by 
\begin{align*}
	c_{\beta\alpha}\simeq -\frac{\lambda_6v^2}{(m^2_{\rho}-m_h^2)}\ll 1\ \to \ c_{\beta\alpha}\simeq-\frac{\lambda_6 v^2}{M^2}.
\end{align*}

\section{Phenomenology}
\label{sec:Pheno}

In this section we discuss the phenomenological signatures of our model. In Section \ref{ssec:PhenononLFV} we briefly review the experimental constraints on $L_e-L_\mu$ gauge interactions, before focussing on the LFV phenomenology in Section \ref{ssec:LFV}. In Section \ref{ssec:neutrino} we address neutrino mixing and the generation of neutrino masses.

\subsection{$L_e-L_\mu$ Gauge Interactions}
\label{ssec:PhenononLFV}

Several experiments search for $Z'$ bosons that interact with SM particles.  In the absence of tree-level kinetic mixing, these searches directly probe the size of the gauge coupling $g'$. Since in our model the $Z'$ interactions arise from gauging the lepton flavour difference $L_e-L_\mu$, $g'$ is primarily constrained by bounds on $Z'ee$ and $Z'\nu\nu$ \cite{Wise:2018rnb,Bauer:2018onh}.\\

\begin{figure}[t]
    \centering
    \includegraphics[width=\linewidth]{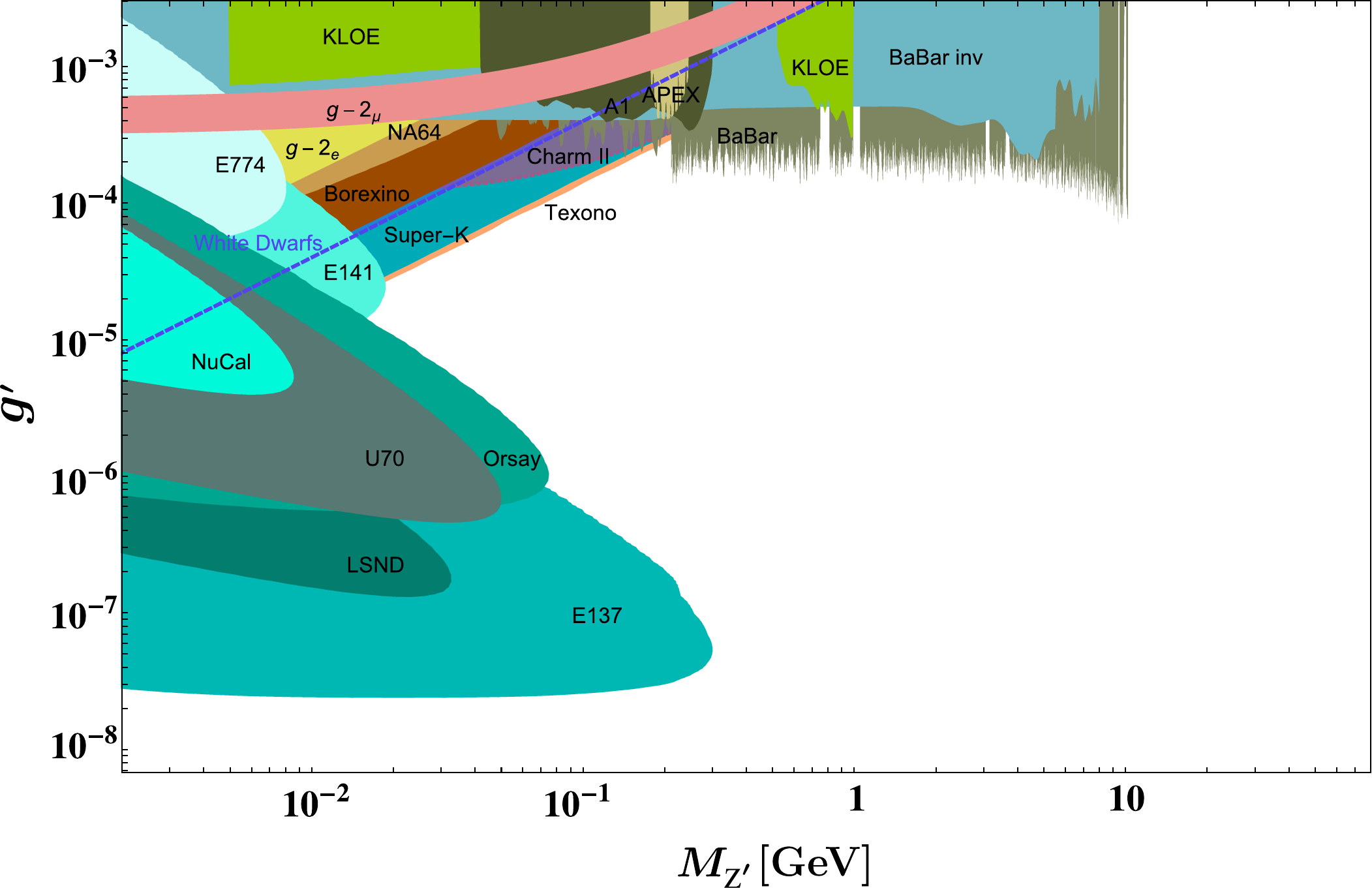}
    \caption{Figure taken from Ref.~\cite{Bauer:2018onh} showing the current bounds on the space of $M_{Z'}$ vs. $g'$ in the case of gauged $L_e-L_\mu$. As described in the text, we consider three benchmark scenarios, one of which involves a $Z'$ boson with a mass $M_{Z'}\sim 10~$GeV and a coupling $g'\sim 10^{-4}$ and is in reach of Belle-II, another benchmark with $M_{Z'}\sim 15~$GeV, which avoids the bounds from BaBar~\cite{BaBar1,BaBar2}, allowing for a gauge coupling of $g'\sim 2.5\times 10^{-3}$ and finally the case of a long-lived $Z'$ with a mass $\text{MeV}\lesssim M_{Z'}<m_\mu-m_e$ and a gauge coupling of $g'\sim 10^{-8}$ which decays outside the Belle-II detector.}
    \label{fig:current_gp_MZp_bounds}
\end{figure}

In the $Z'$ mass range of $10\ \mathrm{MeV} \lesssim M_{Z'}\lesssim 1 $ \GeV, the strongest constraints on the $U(1)'$ gauge coupling come from electron beam dump experiments~\cite{ElectronSearches1,electronSearches2}, neutrino oscillation experiments~\cite{Wise:2018rnb} and neutrino scattering experiments~\cite{neutrinoscattering1,neutrinoscattering2}, while cosmological and astrophysical limits are more relevant for lighter $Z'$ bosons~\cite{Kamada:2015era,Rrapaj:2015wgs,Knapen:2017xzo}. 
Muonium spectroscopy can also probe for $Z'$ bosons in the sub-MeV mass range \cite{Aiba:2021bxe}.
For larger masses, $M_{Z'}\gtrsim 1 $ \GeV, colliders are the most sensitive probes of a fifth force. Below $10$~\GeV, BaBar \cite{BaBar1,BaBar2} set an upper limit of  $g'\lesssim 10^{-4}$ by searching for $Z'$ production in combination with a single photon, $e^+e^-\to \gamma Z'\to \gamma \ell^+\ell^-$. Belle II is expected to push this limit down to $g'\sim 10^{-5}$ \cite{Belle-II:2010dht,Inguglia:2016acz}. $Z'$ bosons with masses beyond $M_{Z'}\sim 10 $ \GeV~can be produced in the decay of an excited meson produced at the LHC, however the process occurs via loop induced kinetic mixing and the bound on the $U(1)'$ gauge coupling is only $g'\lesssim 10^{-1}-10^{-2}$ \cite{LHCb:2017trq,Ilten:2015hya}.

\subsection{Lepton Flavour Violation}
\label{ssec:LFV}

\subsubsection{$\tau \to \mu$ and $\tau\to e$ Transitions}
\begin{figure}
\centering
\includegraphics[scale=1]{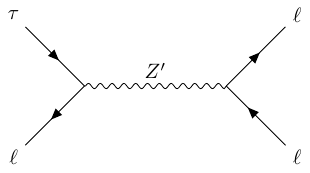}
\caption{$Z'$ mediated $\tau\to 3\ell$ decay, with $\ell=e,\mu$. If $2m_\ell<M_{Z'}<m_\tau-m_\ell$, the decay can happen via the on-shell production of the $Z'$ and its subsequent decay into a lepton pair.}\label{fig:taudecay}
\end{figure}
Via the lepton flavour changing $Z'$ couplings in Eq.~(\ref{eq:ZprimeLFV}), $\tau$ leptons can decay into final states such as $3 e$, $3\mu$, $\mu 2e$ or $e 2\mu$. In Figure \ref{fig:taudecay}, we show the diagram for the tree-level $Z'$ mediated decay of $\tau$ into $3\ell$, where $\ell=e,\mu$. For a $Z'$ lighter than the mass difference $m_\tau-m_\ell$, $\tau$ leptons can decay into an on-shell $Z'$ and a charged lepton $\ell$. If in addition $M_{Z'}>2m_\ell$, the $Z'$ boson can decay into a lepton pair $\ell^+\ell^-$. If this happens before the $Z'$ escapes the detector, a three-lepton final state may be measured as a consequence of the decay chain $\tau\to Z' \ell\to \ell^+ \ell^-\ell$. In the narrow-width approximation, $\mathrm{\Gamma}/M_{Z'}\lesssim g'^2\ll 1$, the Breit-Wigner distribution can be approximated by a $\delta-$function, resulting in a factorized rate   
\begin{align*}
	\mathrm{\Gamma}(\tau\to 3\ell)\simeq \mathrm{\Gamma}(\tau\to Z'\ell)\times \mathrm{Br}(Z'\to \ell^{+} \ell^{-})\,.
\end{align*}
If the muon and electron masses can be neglected with respect to the $Z'$ mass, we can take $\mathrm{Br}(Z'\to \ell^{+} \ell^{-})\sim 1/3$ and~\footnote{Note that there is no divergence in the limit $M_{Z'}\to 0$, because with $v_S, v_\phi\to 0$, the LFV rotation angles $\theta_{i3}$ also vanish.~\cite{Ibarra:2021xyk}}
\begin{align*}
	\mathrm{\Gamma}(\tau\to Z' \ell)=\!\!\!\sum_{X=L,R}\!\!\!\frac{(g'\theta^X_{i3})^2}{64\pi}\frac{m^3_\tau}{M_{Z'}^2}\!\left(1-\frac{M_{Z'}^{2}}{m^2_\tau}\!\right)^{\!\!2}\!\left(1+2\frac{M^2_{Z'}}{m^2_\tau}\!\right)\,,
\end{align*}
where $i=1,2$ for $\ell=e,\mu$ respectively. (The mixing angles $\theta$ are given in terms of the Lagrangian parameters in Eq.~\eqref{eq:anglesmasslepton}). In the case of on-shell $Z'$ production, the rate of $\mathrm{\Gamma}(\tau\to 3\ell)$ scales with $g'^2$ rather than with the $g'^4$ that one would naively expect from the tree-level $Z'$ exchange shown in Figure~\ref{fig:taudecay}. Consequently, the bounds on the flavour off-diagonal Yukawas from $\mathrm{Br}(\tau\to 3e)<2.7\times 10^{-8}$, $\mathrm{Br}(\tau\to 3\mu)<2.1\times 10^{-8}$ are stringent enough to suppress all other LFV signals. Indeed, taking $M_{Z'}=1~$\GeV\  and $g'=10^{-4}$, the upper limit $\mathrm{Br}(\tau\to 3\mu)<2.1\times 10^{-8} $ implies $ y_{23}<3\times 10^{-7} (\sin \beta)^{-1}$. As will become apparent in the following section, $\mu\to e$ processes cannot further constrain the model in this case.

If the $Z'$ is sufficiently long-lived to escape the detector, upper-limits on the off-diagonal Yukawas can be inferred from the lepton-flavour violating decays $\tau\to \ell +{\rm invisible}$. The $Z'$ decay length exceeds $\sim 1$ m when 
\begin{equation}
    (g'M_{Z'})^2\lesssim 2\times 10^{-15}\ {\rm GeV}^2. \label{eq:invisiblecond}
\end{equation}
 In the range $0.1\ {\rm GeV}\lesssim M_{Z'}\lesssim 1.6\ {\rm GeV}$, the extremal $g'$ values compatible with Eq.~(\ref{eq:invisiblecond}) are $g'(0.1\ {\rm GeV})\sim 5\times 10^{-7}$ and $g'(1.6\ {\rm GeV})\sim 3\times 10^{-8}$. Note that these values are not excluded, see Fig.~(\ref{fig:current_gp_MZp_bounds}). Considering $M_{Z'}=0.1$ GeV, the upper limit $\mathrm{Br}(\tau\to \ell+{\rm inv.})/\mathrm{Br}(\tau\to \ell \nu\nu)\lesssim 10^{-3}$ \cite{Belle-II:2022heu} implies
\begin{equation}
    \theta_{i3}^X\lesssim 3\times 10^{-2}\,, \quad i=1,2\,,\quad X=L,R\,.
\end{equation}

 For $0.01\ {\rm GeV}\lesssim M_{Z'}\lesssim 0.1$ GeV, the allowed couplings for $Z'$ bosons that escape the detector are $g'\lesssim 10^{-8}$ and the constraint on the mixing angle arising from $\tau\to\ell+{\rm inv.}$ is $\theta_{i3}\lesssim 0.1$. Smaller $Z'$ masses are tightly constrained by BBN \cite{Ahlgren:2013wba}.
 
If $M_{Z'}>m_\tau$, the $\tau\to 3\ell$ decay is mediated at tree-level by an off-shell $Z'$ boson . Even for a $Z'$ mass of $M_{Z'}\sim 5$~\GeV, treating the $Z'$ exchange as a contact interaction between four leptons is sufficient for an estimation of the decay rate to a $\sim 10\%$ accuracy. Integrating out the $Z'$ boson at the scale of $M_{Z'}$, the four-lepton operators 
\begin{equation}
    \mathcal{O}_{V,XY}=(\bar{\ell} \gamma^\alpha P_X \tau)(\bar{\ell} \gamma_\alpha P_Y\ell) 
    \label{eq:vectorEFT}
\end{equation}
are generated with the following coefficients
\begin{align*}
    C^{e \tau e e}_{V,XY}=-\frac{g'^2\theta^X_{13}}{M^2_{Z'}}\qquad C^{\mu\tau \mu \mu}_{V,XY}=\frac{g'^2\theta^X_{23}}{M^2_{Z'}}.
\end{align*}
The decay rate of $\tau\to 3\ell$ is then given by
\begin{align}
{\rm Br}\left(\tau \to 3\ell\right)=\frac{m_\tau^5 }{1536 \,\pi ^3\, \mathrm{\Gamma}_\tau}
&\left(2 \left|C^{\ell\tau\ell\ell}_{V,LL}\right|^2+\left|C^{\ell\tau\ell\ell}_{V,LR}\right|^2\notag\right.\\
&\left.\quad+L\leftrightarrow R\right)\,,
\label{lto3l}
\end{align}
where $\mathrm{\Gamma}_\tau$ is the total decay width of the $\tau$. Here we are neglecting the QED running from $M_{Z'}$ to the $\tau$ mass. 

\begin{figure}
    \centering
    \includegraphics[width=.8\linewidth]{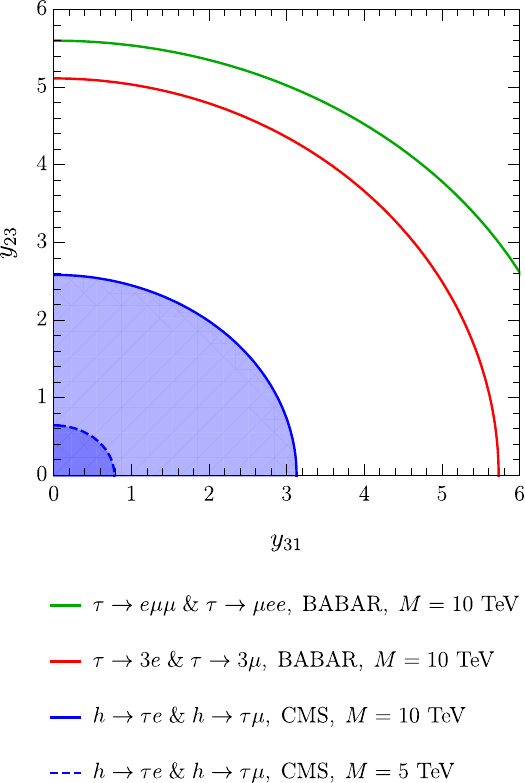}
    \caption{Considering $g'=10^{-4}$, $M_{Z'}=10~$GeV, $t_\beta=\frac{1}{10}$ and $M=10~$TeV, we plot the values of the off-diagonal couplings that saturate the current upper limits on the $\tau$ LFV branching ratios. Since the limit of small mixing (see Eq.~(\ref{eq:anglesmasslepton})) assumes much smaller Yukawas, we conclude that lepton flavour violation in the $\tau$ sector cannot constrain the off-diagonal couplings $y_{31}$ and $y_{23}$ in this region  of parameter space.}
    \label{fig:tauplot}
\end{figure}

Contributions from the scalar sector arise from diagrams similar to the one depicted in Figure \ref{fig:taudecay}, but with a neutral scalar exchange. However, the flavour-diagonal current couples to scalars via the light lepton Yukawas, and LFV Higgs decays are more sensitive to $\tau\leftrightarrow \ell$ flavour changing Yukawa couplings than the Yukawa suppressed $\tau\to 3\ell$ decay \cite{AtlasHiggs}. The width for the LFV Higgs decay is given by~\cite{2HDMZupan}
\begin{align*}
    \mathrm{\Gamma}(h\to \tau \ell)= \frac{\abs{\mathcal{Y}_{\ell\tau}}^2+\abs{\mathcal{Y}_{\tau \ell}}^2}{16\pi}c_{\beta\alpha}^2 m_h
\end{align*}
with $\mathcal{Y}$ as defined in Eq. (\ref{eq:scalarYuk}). The allowed region in the $y_{31}-y_{23}$ plane is plotted in Figure \ref{fig:tauplot}, which shows that $\mathrm{Br}(\tau\to 3e)<2.7\times 10^{-8}$, $\mathrm{Br}(\tau\to 3\mu)<2.1\times 10^{-8}$ and $\mathrm{Br}(h\to \tau\mu)<1.5\times 10^{-3}, \mathrm{Br}(h\to \tau e)<2.2\times 10^{-3}$ \cite{AtlasHiggs}  are all compatible with perturbative Yukawa couplings. These bounds are not able to constrain $y_{31}$ and $y_{23}$ to values smaller than $y_{33}/t_\beta \sim 1/10$, as required in Eq.~\eqref{eq:anglesmasslepton}.

\subsubsection{$\mu\to e$ Transitions}
We now turn to the $\mu \to e$ phenomenology. In presence of the flavour mixing angles of Eq.~\eqref{eq:anglesmasslepton}, the $Z'$ boson can couple at tree-level to $\mu\to e$ currents, however these couplings only arise at second order in the mixing angles and are suppressed by the electron and muon Yukawas: the left-handed and right-handed vector currents come with the products of mixing angles $\theta^L_{13}\theta^L_{23}$ and $\theta^R_{13}\theta^R_{23}$ respectively,
where
\begin{align}
		\theta_{23}^L&= \tan \beta\frac{y_{23}}{y_{33}}\qquad { \theta^R_{23}=\theta_{23}^L\times\frac{y_{22}}{y_{33}}}\nonumber\\
		\theta_{13}^R&= \tan \beta\frac{y_{3 1}}{y_{33}}\qquad {\theta^L_{13}=\theta_{13}^R\times\frac{y_{11}}{y_{33}}}\,. \label{eq:anglestanblepton}
\end{align}
If the $Z'$ is light enough to permit the decay $\mu\to e Z'$, the experimentally allowed values for the gauge coupling are $g'\lesssim 10^{-8}$. Although in the mass range $2m_e\lesssim M_{Z'}\lesssim m_{\mu}-m_e$ the $Z'$ could subsequently decay into an electron positron pair, the decay is not fast enough to happen inside the SINDRUM detector, such that the bound $\mathrm{Br}(\mu\to 3e)<10^{-12}$~\cite{SINDRUM:1987nra} does not apply to the decay chain $\mu\to e (Z'\to \bar{e} e)$. On the other hand,  the non-observation of $\mu\to e+{\rm inv.}$ by the TWIST collaboration sets an upper limit $\mathrm{Br}(\mu\to e Z')<8.1\times 10^{-6}$ \cite{TWIST:2014ymv} in this region of the parameter space. The $Z'$ primarly couples to a right-handed $\mu\to e$ current, while the left-handed couplings are suppressed by the electron Yukawa. The TWIST upper bound $\mathrm{Br}(\mu\to e Z')<8.1\times 10^{-6}$ on the $\mu\to e Z'$ rate
\begin{align*}
	\mathrm{\Gamma}(\mu\to Z' e)=\frac{\left(g'\theta^R_{13}\theta^L_{23}\frac{y_{22}}{y_{33}}\right)^2}{64\pi}\frac{m^3_\mu}{M_{Z'}^2}\!\left(1-\frac{M_{Z'}^{2}}{m^2_\mu}\!\right)^{\!\!2}\!\left(1+2\frac{M^2_{Z'}}{m^2_\mu}\!\right)\,,
\end{align*}
 implies $\theta^R_{13}\theta^L_{23}\lesssim 10^{-3}$, taking  $g'\sim 10^{-8}, M_{Z'}\sim 1$ MeV. This is beyond the reach of $\tau \to e+\text{inv.} $ and $ \tau \to \mu+\text{inv.}$  combined.
 
\begin{figure}
\centering
\subfloat[][]{%
\includegraphics[width=.25\textwidth]{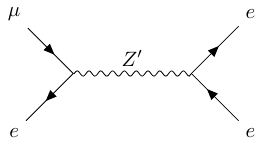}\label{subfig:mu3e}
}
\subfloat[][]{%
\includegraphics[width=.25\textwidth]{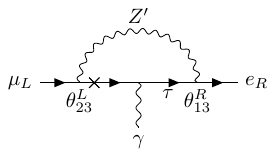}\label{subfig:mutoegamma}
}
\caption{(a) Tree-level $Z'$ exchange contributing to $\mu\to 3e$. (b) $Z'$ penguin diagram contributing to $\mu\to e\gamma$.}
\end{figure}

\begin{figure}[!h]
\centering
\subfloat[][]{%
\includegraphics[scale=1.05]{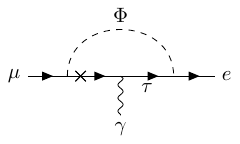}\label{subfig:1loopmutautaue}
}
\qquad
\subfloat[][]{%
\includegraphics[scale=1.05]{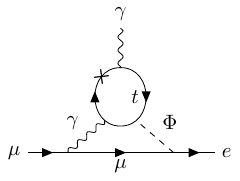}\label{subfig:2looptop}
}
\hfill
\subfloat[][]{%
\includegraphics[scale=1.05]{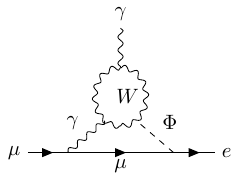}\label{subfig:2loopW}
}
\caption{Loop diagrams with neutral scalars contributing to the rate of $\mu\to e \gamma$. $\Phi=h,\rho, A$}\label{fig:scalarmutoe}
\end{figure}

 For $M_{Z'}>m_\mu$, the $\mu\to 3e$ decay, upon which the bound $\mathrm{Br}(\mu\to 3e)<10^{-12}$ applies, receives contributions from the diagram of Figure \ref{subfig:mu3e}. 
 The effective four-fermion interactions that result from integrating out the $Z'$ boson, are given by
\begin{equation}
    C^{e\mu e e}_{V,XY}=-\frac{g'^2\theta^X_{13}\theta^X_{23}}{M^2_{Z'}}\,. \label{eq:mutoevectors}
\end{equation}
Contributions of the dipole, which comes with the coupling product $\theta^R_{13}\theta^L_{23}$ and avoids the suppression by the SM electron and muon Yukawas, can also be relevant.
% \begin{figure}
% \centering
% \begin{subfigure}{0.4\textwidth}
%     \centering
%     \includegraphics[scale=1]{muto3e.pdf}
%     \caption{}
%     \label{subfig:mu3e}
% \end{subfigure}
% \begin{subfigure}{0.4\textwidth}
%     \centering
% \includegraphics[scale=1]{Pengmutoegamma.pdf}
% \caption{}
% \label{subfig:mutoegamma}
% \end{subfigure}
% \caption{(a) Tree-level $Z'$ exchange contributing to $\mu\to 3e$. (b) $Z'$ penguin diagram contributing to $\mu\to e\gamma$.}
% \end{figure}
For instance, the penguin diagram of Figure \ref{subfig:mutoegamma} contributes to the $\mu\to e$ photon dipole 
\begin{equation}
    \mathcal{O}^{e\mu}_{D,X}=m_\mu(\bar{e}\sigma_{\alpha\beta}P_X \mu)F^{\alpha\beta},
\end{equation}
with a coefficient 
\begin{equation}
    (C^{e\mu}_{D,L})_{peng}=-\frac{3 e}{48\pi^2 M^2_{Z'}}\left(\frac{m_\tau}{m_\mu}\right)g'^2\theta^R_{13}\theta^L_{23}.
\end{equation}
In addition to the penguin, loop diagrams with scalars can give sizeable contributions to the dipole.
In the diagrams of Figure \ref{subfig:1loopmutautaue}, the mass insertion flips the chirality in the virtual $\tau$ line. These diagrams contribute to the dipole coefficient as follows~\cite{Chang:1993kw} 
\begin{align*}
    (C^{e\mu}_{D,L})_\mathrm{1-loop}=\hspace{60mm}&\notag\\
    =-\frac{e}{64\pi^2}\left(\frac{m_\tau}{m_\mu}\right)\bigg[\frac{\mathcal{Y}^*_{\mu\tau}\mathcal{Y}^*_{e\tau}}{m^2_h}c^2_{\beta\alpha}\left(2\log(\frac{m_\tau}{m_h})+\frac{3}{2}\right)&\nonumber\\
    +\frac{\mathcal{Y}^*_{\mu\tau}\mathcal{Y}^*_{e\tau}}{m^2_\rho}s^2_{\beta\alpha}\left(2\log(\frac{m_\tau}{m_\rho})+\frac{3}{2}\right)&\nonumber\\
    -\frac{\mathcal{Y}^*_{\mu\tau}\mathcal{Y}^*_{e\tau}}{m^2_A}\left(2\log(\frac{m_\tau}{m_A})+\frac{3}{2}\right)\bigg]&\,,
\end{align*}
with $\mathcal{Y}$ as defined in Eq.~\eqref{eq:scalarYuk}. $C_{D,R}$ is obtained from $C_{D,L}$ by replacing $\mathcal{Y}\leftrightarrow \mathcal{Y}^\dagger$. 
When the loop is closed by the light Higgs $h$, each flavour changing vertex is proportional to $c_{\beta\alpha}$, resulting in a contribution suppressed by four powers of the heavy Higgs mass, $c^2_{\beta\alpha}\sim1/M^4$. Moreover, in the decoupling limit, the contributions of the heavy scalar $\rho$ and the pseudoscalar $A$ cancel in the above equation, since their masses are degenerate at leading order in $1/M^2$. The total amplitude ends up being a sub-dominant $1/M^4$ effect.
% \begin{figure}[!h]
% \centering
% \begin{subfigure}{0.2\textwidth}
% \centering
% \includegraphics[scale=1]{Scalar1Loop.pdf}
% \caption{}\label{subfig:1loopmutautaue}
% \end{subfigure}
% \\[-10mm]
% \begin{subfigure}{0.2\textwidth}
% \centering
% \includegraphics[scale=1]{BarZee1.pdf}
% \caption{}\label{subfig:2looptop}
% \end{subfigure}
% \hspace{5mm}
% \begin{subfigure}{0.2\textwidth}
% \centering
% \includegraphics[scale=1]{Barzee2.pdf}
% \caption{}\label{subfig:2loopW}
% \end{subfigure}
% \caption{Loop diagrams with neutral scalars contributing to the rate of $\mu\to e \gamma$. $\Phi=h,\rho, A$}\label{fig:scalarmutoe}
% \end{figure}

\begin{figure}[!th]
    \centering
    \includegraphics[width=.8\linewidth]{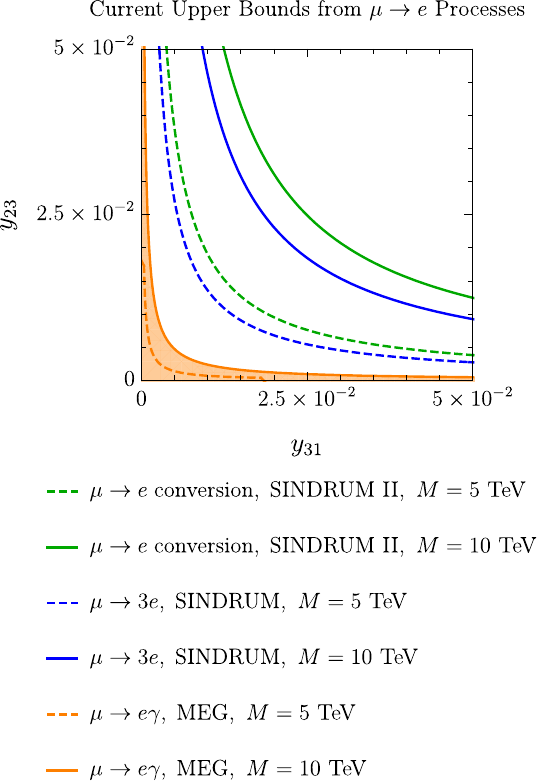}
    \caption{Constraints on the Yukawa couplings $y_{31}$ and $y_{23}$ from experimental upper bounds on $\mu\to e$ transitions. The orange region corresponds to the allowed parameter space. In the plot we consider $g'=10^{-4}$, $M_{Z'}=10$ GeV, $M=10$ TeV (solid lines) or $M=5$ TeV (dashed lines), $\lambda_6=1$ and $\tan\beta=0.1$. The dashed curves show how the lepton flavour violation induced by the scalar sector changes with $M$.  These contributions arise at mass dimension 6 within the SMEFT.}
    \label{fig:mu-e_plot}
\end{figure}

The misalignment of the lepton mass eigenbasis and the lepton gauge eigenbasis also affects the scalar couplings. Whereas the $Z'$ boson acquires $\mu\leftrightarrow e$ couplings proportional to $\propto \theta^L_{13} \theta^L_{23}$ or $\propto \theta^R_{13} \theta^R_{23}$, the scalars acquire $\mu\to e$ couplings proportional to $\theta^R_{13} \theta^L_{23}$. \\
It is well known that the Barr-Zee diagrams of Figures \ref{subfig:2looptop}-\ref{subfig:2loopW} give the leading contribution to $\mu\to e \gamma$ in 2HDM models with $\mu\to e$ Yukawa couplings \cite{PhysRevLett.38.622}. The branching ratio of $\mu\to e \gamma$ is given by
\begin{equation}
    \mathrm{Br}(\mu\to e \gamma)=\frac{m_\mu^5}{4\pi \mathrm{\Gamma}_{\mu}}\left(\abs{C^{e\mu}_{D,L}}^2+\abs{C^{e\mu}_{D,R}}^2\right)\,,
\end{equation}
where $\mathrm{\Gamma}_{\mu}\simeq G_F^2m_\mu^5/(192\pi^3)$ is the total decay rate of the muon, and, to a good approximation,
\begin{equation}
    C^{e\mu}_{D,X}=(C^{e\mu}_{D,X})_\mathrm{peng}+(C^{e\mu}_{D,X})_\mathrm{t-loop}+(C^{e\mu}_{D,X})_\mathrm{W-loop} 
    \label{eq:fulldipole}
\end{equation}
where
$(C^{e\mu}_{D,X})_\mathrm{t-loop}$ and $(C^{e\mu}_{D,X})_\mathrm{W-loop}$ correspond to the two-loop dipole contributions of Figures \ref{subfig:2looptop} and \ref{subfig:2loopW}, respectively, and are defined in Appendix \ref{appendix:BarZee}. Note that these diagrams scale like $1/M^2$ (see also Figure~\ref{fig:mu-e_plot}).

For the $\mu\to 3e$ rate we consider the tree-level diagram of Figure \ref{subfig:mu3e} and the dipole contributions arising from attaching an electron current to the photon. We find 
\begin{align*}
    {\rm Br}&\left(\mu\to 3e\right)=\notag\\
    &-\frac{e^2\, m_\mu^5 }{192\, \pi ^3\,\mathrm{\Gamma}_\mu}\left(|C_{D,L}^{e\mu}|^2+|C_{D,R}^{e\mu}|^2\right) \left(4 \log \left(\frac{m_e^2}{m_\mu^2}\right)+11\right)\notag\\
&+\frac{m_\mu^5 }{1536 \,\pi ^3\, \mathrm{\Gamma}_\mu}
\left(2 \left|C^{e\mu ee}_{V,LL}\right|^2+\left|C^{e\mu ee}_{V,LR}\right|^2
+L\leftrightarrow R\right)\\
&+\frac{e\, m_\mu^5}{192\, \pi ^3 \,\mathrm{\Gamma}_\mu}
\Big({\rm Re}\left[C_{D,R}^{e\mu*} \left( 2\, C^{e\mu ee}_{V,LL}+C^{e\mu ee}_{V,LR}\right)\right]+L\leftrightarrow R\Big)\,,
\end{align*}
with $C^{e\mu ee}_{V,X}$ as defined in Eq.~(\ref{eq:mutoevectors}). 

\begin{figure}
\centering
\includegraphics[scale=1]{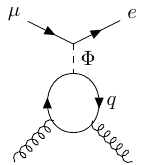}
\caption{Contributions via heavy quark loops to the gluon operator $\mathcal{O}_{GG,X}=(\bar{e}P_X \mu)G^{\alpha\beta}G_{\alpha\beta}$ that enters the $\mu\to e$ conversion rate. $q=c,b,t$ and $\Phi=h,\rho, A$. }\label{fig:gluons}
\end{figure}

LFV interactions with quarks via $Z'$ exchange arise from loop induced kinetic mixing or via penguin diagrams with an external quark current. 
The kinetic mixing is both loop-suppressed and suppressed by the Yukawa couplings featuring in the $Z'\mu e $ vertex, leading to a sub-dominant contribution. 
The relevant scalar contributions are the tree-level exchanges with the $u$, $d$ and $s$ quarks in the nuclei, the Barr-Zee dipoles that already featured in Eq.~\eqref{eq:fulldipole} and are discussed in Appendix~\ref{appendix:BarZee}, as well as the diagrams in Figure~\ref{fig:gluons}. The latter contribute to the gluon operator $\mathcal{O}_{GG,X}=(\bar{e}P_X \mu)G^{\alpha\beta}G_{\alpha\beta}$, which enters the $\mu\to e$ conversion rate. In order to compute these contributions, we integrate out the heavy scalar doublet $H_2$ at the mass scale $M$ and match onto the SMEFT operators $\mathcal{O}_{Ledq}^{e\mu ii}\equiv (\bar L_e \mu)(\bar d_i q_i)$, $\mathcal{O}_{Lequ}^{(1)e \mu ii}\equiv (\bar L_e \mu)(\bar q_i u_i)$, and $\mathcal{O}_{eH}^{e\mu}\equiv (H_1^\dagger H_1)(\bar L_e \mu H_1) $, which leads to the off-diagonal Yukawa couplings of the SM Higgs boson $h$, given in Eq.~\eqref{eq:scalarYuk}. We include the QCD running of the scalar operators from the scale $M$ down to the electroweak scale. At the electroweak scale, the SMEFT operators are matched at tree-level onto the Low Energy Effective Field Theory (LEFT) scalar contact interactions and the gluon operator $\mathcal{O}_{GG,X}=(\bar{e}P_X \mu)G^{\alpha\beta}G_{\alpha\beta}$~\cite{Davidson:2020ord}, which receives contributions from the diagrams with $t$-loops. Finally, we compute the $\mu\to e$ conversion rate using Eq.~(2.22) of \cite{Davidson:2020hkf}, which is written in terms of LEFT operators evaluated at the electroweak scale and includes the running of the Wilson coefficients down to the experiment.

\begin{figure}[!th]
    \centering
    \includegraphics[scale=.8]{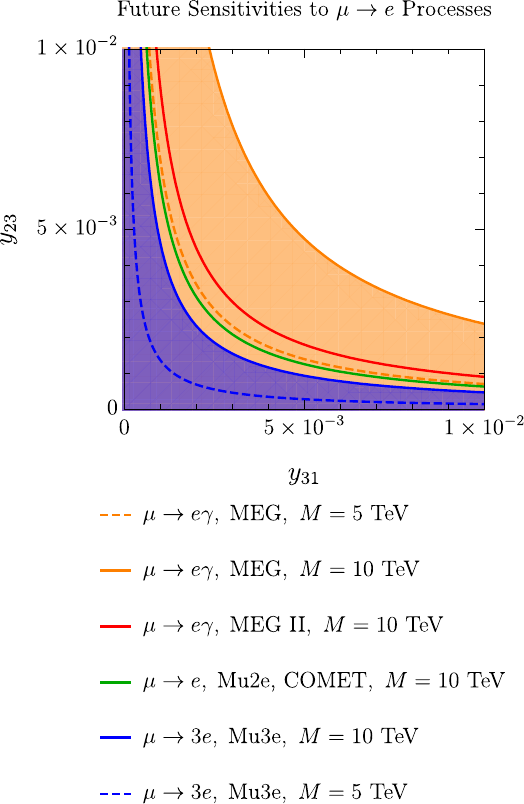}
    \caption{Constraints on the Yukawa couplings $y_{31}$ and $y_{23}$, derived from future experimental sensitivities to $\mu\to e$ transitions.  The light orange region corresponds to the parameter space that will be probed by Mu3e, whereas the purple region is the allowed parameter space if no $\mu\to 3e$ signal is observed by Mu3e. For this plot we consider: $g'=10^{-4}$, $M_{Z'}=10$ GeV, $M=10$ TeV (solid lines) or $M=5$ TeV (dashed lines), $\lambda_6=1$ and $\tan\beta=0.1$.  Again, the suppression of the scalar contributions by (at least) $1/M^2$ can be observed by comparing the solid and the dashed curves.}
    \label{fig:future_mu-e_plot}
\end{figure}

As shown in Figure \ref{fig:mu-e_plot}, the MEG bound, $\mathrm{Br}(\mu\to e\gamma)<4.2\times 10^{-13}$, is better at constraining the product of Yukawa couplings $y_{23}\times y_{31}$ than the current upper limits $\mathrm{Br}(\mu\to 3e)<10^{-12}$ and $\mathrm{Br}(\mu A\to eA)<7\times 10^{-13}$.
On the other hand, future $\mu\to e$ conversion and $\mu\to 3e$ experiments expect an impressive improvement in the branching ratio sensitivities. The COMET and Mu2e collaborations aim at $\mathrm{Cr}(\mu \mathrm{Al}\to e \mathrm{Al})\sim 10^{-16}$ \cite{COMET:2009qeh,Mu2e:2014fns} and Mu3e at $\mathrm{Br}(\mu\to 3e)\sim 10^{-16}$ \cite{Blondel:2013ia}, both of which can surpass the expected sensitivity of MEG-II with $\mathrm{Br}(\mu\to e \gamma)\sim 6\times 10^{-14}$ \cite{MEGII:2018kmf} (Figure \ref{fig:future_mu-e_plot}).

A $Z'$ boson with $M_{Z'}=10$ GeV and $g'=10^{-4}$, as considered in Figures \ref{fig:mu-e_plot} and \ref{fig:future_mu-e_plot},  will be probed at Belle-II in the upcoming years. In this scenario $\mu\to e$ rates are the result of contributions both from the $Z'$ and from scalar LFV interactions. The scalar contributions dominate over the $Z'$ contributions in this region of parameter space.

\begin{figure}
    \centering
    \includegraphics[scale=.8]{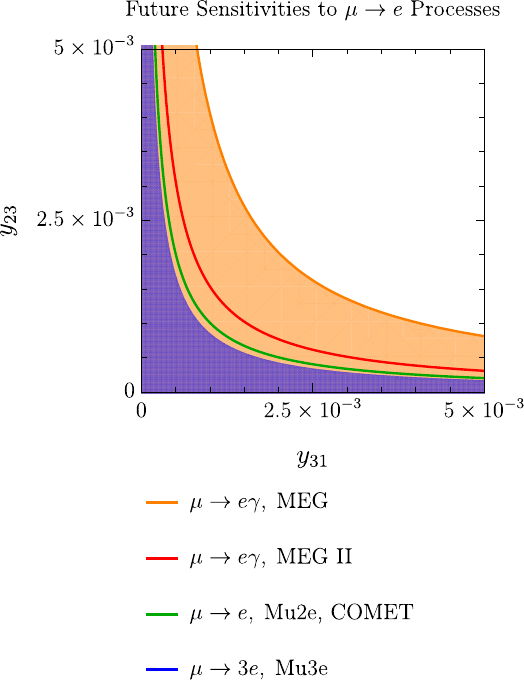}
    \caption{Future sensitivities of searches for $\mu\to e$ processes to the Yukawa couplings $y_{31}$ and $y_{23}$ for $g'=2.5\times 10^{-3}$, $M_{Z'}=15$~GeV and $\tan\beta=0.1$. Since in this case the contributions of the scalar singlet cannot be neglected with respect to those of the scalar doublet, we only show the vector contributions (which are independent of $M$ and $\lambda_6$).} 
    \label{fig:futuremueplotMZ15}
\end{figure}
By considering a marginally larger mass, $M_{Z'}=15$~\GeV, the BaBar constraint $g\lesssim 10^{-4}$ can be avoided and the most stringent upper limit on $g'$ becomes $g'\lesssim 10^{-1}-10^{-2}$ from LHCb \cite{Ilten:2015hya,LHCb:2017trq}. Taking $g'\sim 2.5\times 10^{-3}$ and $M_{Z'}\sim 15$ \GeV, the singlet has a mass $M_S\simeq v_s\simeq M_{Z'}/g'\sim 10\ \mathrm{TeV}$, which is of the same order as the heavy scalar masses of the 2HDM sector, and thus cannot be integrated out. In Figure~\ref{fig:futuremueplotMZ15} we show the $\mu\to e$ sensitivities in the $y_{31}-y_{23}$ plane considering only vector contributions and $g'=2.5\times 10^{-3}$, $M_{Z'}=15\ \GeV$, while the inclusion of the scalar diagrams would require a more careful analysis of the singlet-doublet mixing.
\\
\\

In summary, we observe regions of parameter space in which $\mu\to e=\mu\to \tau \times \tau\to e$ can constrain the model more than direct searches for $\tau\to \mu$ and $\tau\to e$:
\begin{itemize}
    \item When the decay $\mu\to e Z'$ is kinematically allowed, $\mu\to e+\ {\rm inv.}$ searches can probe smaller off-diagonal Yukawa couplings than $\tau\to \ell+\ {\rm inv.}$. 
    \item For $M_{Z}\gtrsim m_\tau$, $\tau$ LFV transitions are unable to constrain the model below $y_{31},y_{23}\sim \order{1}$, while $\mu\to e$ leads to the constraints shown in Figs.~(\ref{fig:mu-e_plot})-(\ref{fig:future_mu-e_plot}).
\end{itemize}

\subsection{Neutrino masses}
\label{ssec:neutrino}
In this section we show that our model is compatible with neutrino mass differences and oscillation data. 

After spontaneous symmetry breaking, the neutrino mass Lagrangian takes the form
\begin{align*}
     \mathcal{M}_{\nu}&= (M_D)_{ij}\bar{L}_i  N_j+\frac{1}{2}\overline{N^c_i}(M_N)_{ij} N_j+\mathrm{h.c.}\,,
\end{align*}
where
\begin{align*}
    M_D=\begin{pmatrix}
    m^\nu_{11} & & m^\nu_{13}\\
     & m^\nu_{22}& \\
     & m^\nu_{32} & m^\nu_{33}
    \end{pmatrix}\,\qquad     M_N=\begin{pmatrix}
     & M^N_{12} & M^N_{13}\\
     \cdot & & \\
     \cdot & & M^N_{33}
    \end{pmatrix}\,.
\end{align*}
When the Majorana masses $M_N$ are much larger than the Dirac masses $M_D$, the mass matrix of the light active neutrinos is obtained via the celebrated seesaw formula \cite{MINKOWSKI1977421}
\begin{equation}
    M_\nu=-M_D M^{-1}_R M^T_D\,.
\end{equation}
The PMNS matrix $U$ diagonalises the neutrino mass matrix $U^T M_\nu U=\mathrm{diag}\begin{pmatrix}
m_1 & m_2 & m_3
\end{pmatrix}$ and is canonically parameterised as
\begin{footnotesize}
\begin{align*}
	U=&\begin{pmatrix}
		c_{12}c_{13} & s_{12}c_{13} & s_{13}e^{-i\delta}\\
		-s_{12}c_{23}-c_{12}s_{23}s_{13}e^{i\delta} & c_{12}c_{23}-s_{12}s_{23}s_{13}e^{i\delta} & s_{23}c_{13}\\
		s_{12}s_{23}-c_{12}c_{23}s_{13}e^{i\delta} & -c_{12}s_{23}-s_{12}c_{23}s_{13}e^{i\delta} & c_{23}c_{13}
	  \end{pmatrix}\notag\\
	  &\qquad \times \mathrm{diag}\begin{pmatrix}1 & e^{i\alpha_{12}} & e^{i\alpha_{31}}\end{pmatrix}\,.
\end{align*}
\end{footnotesize}
Assuming the so-called Normal Ordering (NO) $m_1<m_2<m_3$, a recent global fit \cite{deSalas:2020pgw} to neutrino oscillation data gives the following values for the mass squared differences and mixing angles:
\begin{align}
	m^2_{2}-m^2_{1}=&\left[6.94-8.14\right]\times 10^{-5}\ e\mathrm{V}^2\notag\\ \abs{m^2_{3}-m^2_{1}}=&\left[2.47-2.63\right]\times 10^{-3}\ e\mathrm{V}^2 \nonumber\\
	\sin^2\theta_{12}=&\left[2.71-3.69\right]\times 10^{-1}\notag\\ \sin^2\theta_{23}=&\left[4.34-6.1\right]\times 10^{-1} \nonumber\\ 
	\sin^2\theta_{13}=&\left[2.000 - 2.405\right]\times 10^{-2}\notag\\
	\delta=&\left[0.71 - 1.99\right]\times \pi \label{eq:NeutrinoFIT}
\end{align}
Here the lower and upper values define the $\pm3\sigma$ range. The texture of the neutrino mass matrix in our model is compatible with the observed mass differences and mixing angles. To simplify the expressions, we consider $m^\nu_{32}\to 0$, resulting in a neutrino mass matrix that reads
\begin{align*}
    M_\nu=&\begin{pmatrix}
    -\frac{(m^\nu_{13})^2}{M^N_{33}} & -\frac{m^\nu_{22}m^\nu_{11}}{M^N_{12}}\left(1-\frac{m^\nu_{13}M^N_{13}}{m^\nu_{11}M^N_{33}}\right) & -\frac{m^\nu_{13}m^\nu_{33}}{M^N_{33}} \\
    \cdot & -\frac{(m^\nu_{22})^2(M^N_{13})^2}{M^N_{33}(M^N_{12})^2} & -\frac{(m^\nu_{22})(m^\nu_{33})(M^N_{13})}{M^N_{33}(M^N_{12})^2} \\
    \cdot & \cdot & -\frac{(m^\nu_{33})^2}{M^N_{33}}   \end{pmatrix} \\
    \equiv &
    \begin{pmatrix}
    \frac{a^2}{Z} & X & a \\
    \cdot & Y & b \\
    \cdot & \cdot & Z 
    \end{pmatrix}\,. \nonumber
\end{align*}
We find that with the choice of parameters 
\begin{align*}
    a\sim  2\times 10^{-3}~\mathrm{eV}, \; b\sim -2.75 \times 10^{-2}~\mathrm{eV}\,, \\
    X\sim 9 \times 10^{-3}~\mathrm{eV},\; Y\sim -2 \times 10^{-2}~\mathrm{eV},\\
    Z\sim -2 \times 10^{-2}~\mathrm{eV}\,,
\end{align*}
we are within the parameter ranges listed in Eq.~(\ref{eq:NeutrinoFIT}), and predict the neutrino masses
\begin{align*}
    m_1\sim & 2 \times 10^{-3}~\mathrm{eV},\\
    m_2\sim & 9 \times 10^{-3}~\mathrm{eV},\\
    m_3\sim & 5 \times 10^{-2}~\mathrm{eV}\,.
\end{align*}
Assuming that the sterile neutrinos have masses in the TeV range, which can be probed at the LHC~\cite{CMS:2018iaf,ATLAS:2019kpx}, the above values require Dirac Yukawas that are $y^\nu\sim \order{10^{-7}}$.

The interactions generating neutrino masses in our model cannot lead to detectable charged lepton flavour violating signals.

\section{Conclusions}
%We have proposed a model which extends 
In this article we proposed a simple model that shows how $\mu\to e$ processes can probe $\tau \to e $ and $\tau\to\mu$ couplings beyond the reach of direct searches for lepton flavour violation in the $\tau$ sector.
We extended
the Standard Model gauge group by the anomaly-free abelian group $U(1)'\equiv U(1)_{L_e-L_\mu}$
%. Two scalars that are charged under this $U(1)'$ group are added to the SM particle content, one $SU(2)_L$ doublet, $\phi$, and one singlet, $S$.
and added two scalars that are charged under this $U(1)'$ group, one $SU(2)_L$ doublet, $\phi$, and one singlet, $S$, to the SM particle content.

As a result of the spontaneous breaking of the $U(1)'$ gauge group, the associated $Z'$ boson acquires a mass. Since the new scalar doublet is singly charged under the $U(1)'$ gauge group, $\mu\leftrightarrow \tau$ and $\tau\leftrightarrow e$  Yukawa couplings are allowed, while $\mu \leftrightarrow e$ couplings are forbidden. Nonetheless, $\mu\to e$ can be mediated by the product of $\mu\to \tau\times \tau \to e$ interactions. After electroweak symmetry breaking, the $Z'$ couplings receive flavour off-diagonal components due to the misalignment of the gauge eigenbasis and the mass eigenbasis of the leptons.

If $M_{Z'}\lesssim m_\mu-m_e$ and the $Z'$ boson decays outside the detector, searches for $\mu\to e +\text{inv.}$ can compete with constraints on $\tau\to e$ and $\tau\to \mu$ couplings from $\tau\to e +\text{inv.}$ and $\tau\to \mu +\text{inv.}$ searches.

Also in the case of $M_{Z'}\gtrsim m_\tau$, $\tau\to \ell$ searches do not appreciably constrain the model and the $\mu\to e$ processes lead to the most stringent limits, despite being proportional to the product of Yukawa couplings $y_{23}\times y_{31}$. The model predicts rates that are in reach of the upcoming $\mu\to e$ experiments and our $Z'$ boson can be searched for at Belle-II.

Finally, if we add sterile neutrinos that are charged under $U(1)'$ to the spectrum, the singlet VEV can contribute to their Majorana masses. We show that, via a type-I seesaw mechanism, we can accommodate for the observed neutrino masses and mixing angles. 
\\

Our model shows explicitly that allowing for $\tau \to\mu$ and $\tau \to e$ processes close to the current experimental bounds can lead to observable effects in $\mu\to e$ searches.

\label{sec:Conclusion}

\section*{Acknowledgments}
We thank Sacha Davidson for her useful insights and feedback.
M.A. is supported by a doctoral fellowship from the IN2P3 and thanks the Physik-Institut of the University of Zurich for its hospitality during the completion of this work. 
F.K. acknowledges support by the grant PP00P2\_176884 of the Swiss National Science Foundation and thanks the Laboratoire Univers et Particules de Montpellier for its hospitality. 
Last but not least, we would like to thank Bella for her moral support.

\appendix

\section{Barr-Zee Contributions}\label{appendix:BarZee}
In this appendix, we give the most relevant two-loop Barr-Zee type contributions to the $\mu\to e$ dipole for reference~\cite{Chang:1993kw}. The corresponding diagrams are shown in Figures \ref{subfig:2looptop}-\ref{subfig:2loopW}.\\
The top loop contribution to the dipole is given by
\begin{align*}
    &(C^{e\mu}_{D,L})_\mathrm{t-loop}=\frac{e\alpha}{12\pi^3}\frac{1}{m_\mu m_t}\\
    &\quad\times \left[\sum_{\Phi=h,\rho}F^{e\mu}_{L,\Phi}F^{tt}_{L,\Phi}f\left(\frac{m^2_t}{m^2_\Phi}\right)+F^{e\mu}_{L,A}F^{tt}_{L,A}g\left(\frac{m^2_t}{m^2_\Phi}\right)\right]\,,
\end{align*}
\begin{align*}
    &(C^{e\mu}_{D,L})_\mathrm{t-loop}=\frac{e\alpha}{12\pi^3}\frac{1}{m_\mu m_t} \\
    &\quad\times\left[\sum_{\Phi=h,\rho}F^{e\mu}_{R,\Phi}F^{tt}_{R,\Phi}f\left(\frac{m^2_t}{m^2_\Phi}\right)+F^{e\mu}_{R,A}F^{tt}_{R,A}g\left(\frac{m^2_t}{m^2_\Phi}\right)\right]\,,
\end{align*}
where (substituting $s_{\beta\alpha}\sim 1)$
\begin{align*}
    F^{e\mu}_{L,h}&=\frac{\mathcal{Y}^*_{\mu e}}{\sqrt{2}}c_{\beta\alpha}\quad
    F^{e\mu}_{L,\rho}=-\frac{\mathcal{Y}^*_{\mu e}}{\sqrt{2}}\quad
    F^{e\mu}_{L,A}=-i\frac{\mathcal{Y}^*_{\mu e}}{\sqrt{2}}\\
    F^{e\mu}_{R,h}&=\frac{\mathcal{Y}_{e\mu}}{\sqrt{2}}c_{\beta\alpha}\quad F^{e\mu}_{R,\rho}=-\frac{\mathcal{Y}_{e\mu}}{\sqrt{2}}\quad F^{e\mu}_{R,A}=i\frac{\mathcal{Y}_{e\mu}}{\sqrt{2}}\\ F^{tt}_{X,h}&=\frac{m_t}{v}(1-c_{\beta\alpha}\tan\beta) \quad
    F^{tt}_{X,\rho}=\frac{m_t}{v}(c_{\beta\alpha}+\tan\beta)\\   F^{tt}_{X,A}&=-i\frac{m_t}{v}\tan\beta\,.
\end{align*}
The loop functions $f$ and $g$ are defined as~\cite{Chang:1993kw}
\begin{align*}
    f(z)=&\frac{1}{2} z\int_0^1 dx \frac{1-2x\left(1-x\right)}{x\left(1-x\right)z}\log\frac{x(1-x)}{z}\\
    g(z)=&\frac{1}{2} z \int_0^1 dx \frac{1}{x(1-x)-z}\log\frac{x(1-x)}{z}\,.
\end{align*}
The $W$ loop of Figure \ref{subfig:2loopW} leads to the dipole contribution
\begin{align}
    &(C^{e\mu}_{D,L})_\mathrm{W-loop}=\notag\\
    &=\frac{e\alpha}{32\pi^3}c_{\beta\alpha}\frac{\mathcal{Y}^*_{\mu e}}{\sqrt{2}m_\mu v}\notag\\
    &\times \left( 3f\left(z_h\right)+5g\left(z_h\right)+\frac{3}{4}(g\left(z_h\right)+h\left(z_h\right))+\frac{f\left(z_h\right)-g\left(z_h\right)}{2z_h}\!\!\right)\nonumber\\
    &-\frac{e\alpha}{32\pi^3}c_{\beta\alpha}\frac{\mathcal{Y}^*_{ \mu e}}{\sqrt{2}m_\mu v}\notag\\
    &\times \left( 3f\left(z_\rho\right)+5g\left(z_\rho\right)+\frac{3}{4}(g\left(z_\rho\right)+h\left(z_\rho\right))+\frac{f\left(z_\rho\right)-g\left(z_\rho\right)}{2z_\rho}\right)\nonumber
\end{align}
\begin{align}
    &(C^{e\mu}_{D,R})_\mathrm{W-loop}=\notag\\
    &=\frac{e\alpha}{32\pi^3}c_{\beta\alpha}\frac{\mathcal{Y}_{ e\mu}}{\sqrt{2}m_\mu v}\notag\\
    &\times \left(3f\left(z_h\right)+5g\left(z_h\right)+\frac{3}{4}(g\left(z_h\right)+h\left(z_h\right))+\frac{f\left(z_h\right)-g\left(z_h\right)}{2z_h}\right)\nonumber\\
    &-\frac{e\alpha}{32\pi^3}c_{\beta\alpha}\frac{\mathcal{Y}_{e\mu}}{\sqrt{2}m_\mu v}\notag\\
    &\times \left(3f\left(z_\rho\right)+5g\left(z_\rho\right)+\frac{3}{4}(g\left(z_\rho\right)+h\left(z_\rho\right))+\frac{f\left(z_\rho\right)-g\left(z_\rho\right)}{2z_\rho}\right)\,,\nonumber
\end{align}
where we have defined $z_\Phi=m_W^2/m^2_\Phi$. The loop function $h$ is defined as~\cite{Chang:1993kw}
\begin{align*}
    h(z)=&z^2\frac{\partial}{\partial z}\left(\frac{g(z)}{z}\right)\notag \\
    =& \frac{z}{2}\int_0^1 \frac{dx}{z-x(1-x)} 
    \left(1+ \frac{z}{z-x(1-x)}\log\frac{x(1-x)}{z}\right)\,.
\end{align*}

\bibliography{references.bib}

\end{document}